\documentclass[journal, 10pt]{IEEEtran}
\usepackage{lineno,hyperref}
\usepackage{amsmath,amssymb}

\usepackage{booktabs}
\usepackage{makecell}
\usepackage{multirow}
\usepackage{graphicx}
\usepackage{multirow}
\usepackage{tabu,bm}
\usepackage{color}
\usepackage{subfigure}
\usepackage{lipsum,mwe,cuted}
\usepackage{stfloats}
\hypersetup{hidelinks}

\usepackage{algorithm}  
\usepackage{algorithmic}
\usepackage{url}
\usepackage{cite}
\usepackage{enumitem} 
\usepackage{diagbox} 
\ifCLASSINFOpdf
\else
\fi
\begin{document}
\title{A Lite Distributed Semantic Communication System for Internet of Things}
% Pruning, Quantization and Cluster for distributed
%

\author{Huiqiang~Xie,~\IEEEmembership{Graduate Student Member,~IEEE,} and
        Zhijin~Qin,~\IEEEmembership{Member,~IEEE}
        % <-this % stops a space
\thanks{Huiqiang Xie and Zhijin Qin are with the School of Electronic Engineering and Computer Science, Queen Mary University of London, London E1 4NS, UK (e-mail: h.xie@qmul.ac.uk, z.qin@qmul.ac.uk).}% <-this % stops a space
}

\markboth{Submit to IEEE Journal on Selected Areas in Communications }%
{Shell \MakeLowercase{\textit{et al.}}: Bare Demo of IEEEtran.cls for IEEE Journals}

\maketitle

\begin{abstract}
The rapid development of deep learning (DL) and widespread applications of Internet-of-Things (IoT) have made the devices smarter than before, and enabled them to perform more intelligent tasks. However, it is challenging for any IoT device to train and run DL models independently due to its limited computing capability.  In this paper, we consider an IoT network where the cloud/edge platform performs the DL based semantic communication (DeepSC) model training and updating while IoT devices perform data collection and transmission based on the trained model. To make it affordable for IoT devices, we propose a lite distributed semantic communication system based on DL, named L-DeepSC, for text transmission with low complexity, where the data transmission from the IoT devices to the cloud/edge works at the semantic level to improve transmission efficiency. Particularly, by pruning the model redundancy and lowering the weight resolution, the L-DeepSC becomes affordable for IoT devices and the bandwidth required for model weight transmission between IoT devices and the cloud/edge is reduced significantly. Through analyzing the effects of fading channels in forward-propagation and  back-propagation during the training of L-DeepSC, we develop a channel state information (CSI) aided training processing to decrease the effects of fading channels on transmission. Meanwhile, we tailor the semantic constellation to make it implementable on capacity-limited IoT devices. Simulation demonstrates that the proposed L-DeepSC achieves competitive performance compared with traditional methods, especially in the low signal-to-noise (SNR) region. In particular, while it can reach as large as 40x compression ratio without performance degradation.
\end{abstract}


% \fi
\IEEEpeerreviewmaketitle

% \section{Introduction}

%(a) the communication between cloud platform and IoT devices (b) the communication between IoT devices. 
\section{Introduction}
With the widely deployed connected devices, Internet of Things (IoT) networks are providing more and more intelligent services, i.e., smart home, intelligent manufacturing, and smart cities,  by processing a massive amount of data generated by those connected devices \cite{atzori2010internet, QiuCLAZ18}. Deep learning (DL) \cite{goodfellow2016deep} has demonstrated great potentials in  processing various types of data, i.e., images and texts. The DL-enabled IoT devices are capable of exploiting and processing different types of data more effectively as well as handling more intelligent tasks than before. Although some IoT devices have certain capability to process simple DL models, the limited memory, computing, and battery capability still prevent from wide applications of DL \cite{mohammadi2018deep}. Therefore, the burden of DL model updates is usually transferred to the cloud/edge platform \cite{li2018learning}. Particularly, the DL model is trained at the cloud/edge platform based on data from the IoT devices, and then the trained model is distributed to IoT devices. However, data transmitted over the air could be distorted by wireless channels, which may cause improper trained results, i.e., local optimum. Moreover, the large number of parameters in DL models leads to high latency when distributing the DL models with limited bandwidth. Therefore, transmitting accurate data to the cloud/edge platform over wireless channels for model training and reducing the number of parameters in DL models for lower latency and power consumption at the IoT devices are two crucial problems.

\begin{figure}[tp!]
	\centering
	\includegraphics[width=85mm]{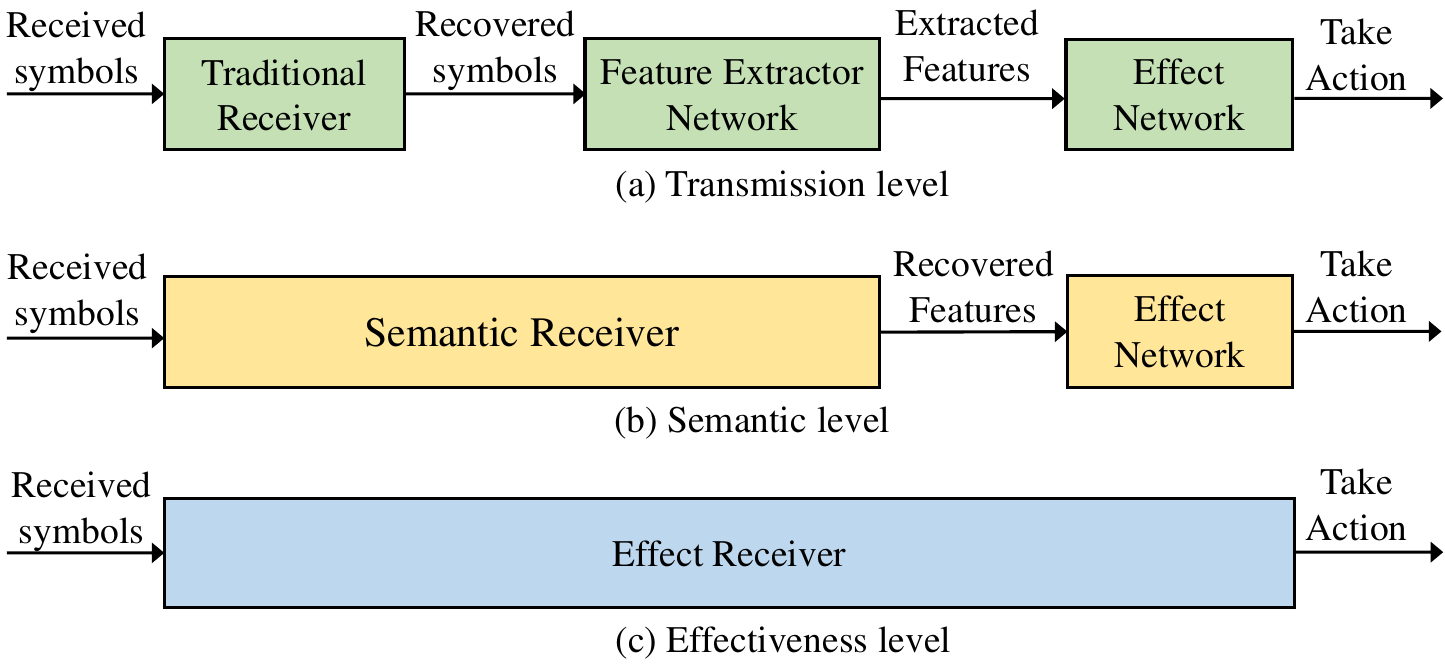}
	\caption{Illustration of three communication levels at the receiver.}
	\label{three levels}
\end{figure}

To address the first problem on accurate data transmission in an IoT network, semantic communication system, which interprets information at the semantic level rather than bit sequences \cite{carnap1952outline}, is promising. To make a decision based on the received information, there are usually three steps: i) the traditional communication receiver to recover the raw data \cite{tse2005fundamentals}; ii) the feature extractor to obtain and interpret the meanings of the raw data for the decision \cite{guyon2008feature}; and iii)  the effects network to produce the desired effects according to the extracted features \cite{szeliski2010computer, indurkhya2010handbook}. Correspondingly, the communication is categorized into three levels \cite{weaver1953recent}, including transmission level to guarantee the transmission accuracy of bit sequence, semantic level to guarantee the exchange  of semantic information, and effectiveness level to measure the corresponding effects or caused actions of transmitted information, i.e., network re-configuration, which is illustrated in Fig. \ref{three levels}. The traditional communication system works at the transmission level shown in Fig. \ref{three levels}(a), which aims at transmitting and receiving symbol accurately \cite{wireless1}. The followed feature extractor network and effect networks are designed separately based on applications. However, designing these modules separately may lead to \textit{error propagation} and prevent from reaching joint optimality. For example, the feature network is not able to correct errors from the traditional receiver, which will affect the subsequent decision making in the effect network. Thus, through designing the traditional receiver and feature extractor network jointly (the semantic level) or merging traditional receiver, feature extractor network, and effects network together (the effectiveness level), communication systems have the capability of error correction at the semantic level and effectiveness level, respectively. In this paper, we will focus on distributed semantic communications for IoT networks and leave effectiveness level communication to future research.

With the recent advancements on DL, it is promising to represent a traditional transceiver or each individual signal processing block by a deep neural network (DNN) \cite{qin2019deep}.  Inspired by the autoencoder in DL techniques, an end-to-end (E2E) communication system has been proposed to merge the signal processing blocks in traditional communication \cite{o2017introduction}. Missing channel gradients becomes the bottleneck of training E2E communication systems. There are several works for mitigating this problem \cite{DornerCHB18, aoudia2019model, ye2020deep}. D{\"{o}}rner \textit{et al.} proposed a two-phase training processing \cite{DornerCHB18} by training the transceiver with a stochastic channel model firstly, and fine-toning the receiver over real channels. Aoudia \textit{et al.} estimated the channel gradients by sampling from a relaxed distribution based on stochastic reinforce learning policy \cite{aoudia2019model}, where the transmitter and receiver are trained separately. Ye \textit{et al.} proposed generative adversarial network (GAN) to approximate the unknown channel model \cite{ye2020deep} so that the channel gradients can be estimated by the GAN.

There have been some initial works related to deep semantic communications \cite{bourtsoulatze2019deep, DBLPLee2019, jankowski2020deep, gold2018, Xie2020deep}. Bourtsoulatze \textit{et al.} \cite{bourtsoulatze2019deep} proposed joint source-channel coding for wireless image transmission  based on the convolutional neural network (CNN), where peak signal-to-noise ratio (PSNR) is used to measure the accuracy of image recovery at the receiver. Taking image classification tasks into consideration,  Lee \textit{et al.} \cite{DBLPLee2019} developed a transmission-recognition communication system by merging wireless image transmission  with the effect network as DNNs, i.e., image classification, which achieves higher image classification accuracy than performing them separately. For texts, Farsad \textit{et al.} \cite{gold2018} designed joint source-channel coding for erasure channel by  using a recurrent neural network (RNN) and a fully-connected neural network (FCN), where the system recovers the text directly rather than perform channel and source decoding separately. In order to understand texts better and adapt to dynamic environments, Xie \textit{et al.} \cite{Xie2020deep} developed  a semantic communication system based on Transformer, named DeepSC, which clarifies the concepts of semantic information and semantic error at the sentence-level for the first time. In brief, compared with traditional approaches, the semantic communication systems are more robust to channel variation and are able to achieve better performance in terms of source recovery and image classification, especially in the low signal-to-noise (SNR) regime.

To deal with the second problem in reducing the number of parameters, network slimmer has attracted extensive attention to compress DL models without degrading performance since neural networks are usually over-sized \cite{denton2014exploiting}. \textit{Parameters pruning} and \textit{quantization} are two main approaches for DL model compression. Parameter pruning is to remove the unnecessary connections between two neurons or important neurons. Han \textit{et al.} \cite{han2015learning} proposed an iterative pruning approach, where the model is trained first, then pruned by a given threshold, and is fine-tuned to recover performance in terms of image classification. This approach could reduce the connections without losing accuracy. Liu \textit{et al.} \cite{LiuLSHYZ17} proposed to prune the filters in CNN by training the model with the $L_1$ regularization so that redundancy weights converge to zero directly without sacrificing the performance. By analyzing the connection  sensitivity among neurons and layers,  Li  \textit{et al.} \cite{0022KDSG17} remove the insensitive layers, which further increases inference speed. By applying these pruning approaches, DL models can be compressed by 13 to 20 times. Quantization aims to represent a weight parameter with lower precision (fewer bits), which reduces the required bitwidth of data flowing through the neural network model in order to shrink the model size for memory saving and simplify the operations for computing acceleration \cite{krishnamoorthi2018quantizing}.  With vector quantization, Gong  \textit{et al.} \cite{gong2014compressing} quantize the DL models. Similarly, Zhou \textit{et al.} \cite{ZhouYGXC17} investigated an iterative quantization, which starts with a trained full-resolution model and then quantizes only a portion of the model followed by several epochs of re-training to recover the accuracy loss from quantization. A mix precision quantization by Li \textit{et al.} \cite{li2016ternary} quantizes weights while keeping the activations at full-resolution. The training algorithm by Jacob \textit{et al.} \cite{jacobKCZTHAK18} preserves the model accuracy after post-quantization. With the quantization, the weights can generally be compressed from 32-bit to 8-bit without performance loss. Similarly, pruning and quantizing can be also used in DL-enabled communication systems. For example, Guo \textit{et al.} \cite{jiajia2020} have shown that model compression can accelerate the processing of channel state information (CSI) acquisition and signal detection in massive multiple-input multiple-output (MIMO) systems without performance degradation.

Through applying network slimmer into our existing work DeepSC, the aforementioned two challenges in IoT networks can be effectively addressed. Although the above works validate the feasibility, we still face the following issues to make it affordable for IoT devices:
\begin{itemize}
\item \textit{Question 1: How to design semantic communication systems over wireless fading channels?}
\item \textit{Question 2: How to form the constellation to make it affordable for capacity-limited IoT devices?}
\item \textit{Question 3: How to compress semantic models for fast-model transmission and low-cost implementation on IoT devices?}
\end{itemize} 

In this paper, we design a distributed semantic communication system for IoT networks. Especially, a lite DeepSC is proposed (L-DeepSC) to address the above questions.  Different from our previous work \cite{Xie2020deep},  this work solves the training DeepSC problem over fading channels with imperfect CSI and considers different wireless channel models to show the generalization of our method. Moreover, this work extends \cite{Xie2020deep} to a more practical IoT scenario, where two key problems, model updating, and broadcasting, are solved. This work also addresses the issue of the finite constellation sizes for capacity-constrained IoT devices while \cite{Xie2020deep} assumes infinite constellations. The main contributions  of this paper are summarized as follows.
\begin{itemize}
    \item We design a distributed semantic communication network under power and latency constraints, in which the receiver and feature extractor networks are jointly optimized by overcoming fading channels.
    \item By identifying the impacts of CSI on DL model training over fading channels, we propose a CSI-aided semantic communication system to speed up convergence, where the CSI is refined by a de-noise neural network. This addresses the aforementioned \textit{Question 1}.
    \item To make data transmission and receiving affordable for capacity-constrained devices, we design a  finite-bits constellation to solve \textit{Question 2}.
    \item Due to over-parametrization, we propose a model compression algorithm, including network sparsification and  quantization, to reduce the size of DL models by pruning the redundancy connections and quantizing the weights,  which addresses the aforementioned \textit{Question 3}.
\end{itemize}

The rest of this paper is organized as follows.  The distributed semantic communication system model is introduced and the corresponding problems are identified in Section II. Section III presents the proposed L-DeepSC. Numerical results are used to verify the performance of the proposed L-DeepSC in Section IV. Finally, Section V concludes this paper.

$Notation$: $\mathbb{C}^{n \times m}$ and $\mathbb{R}^{n \times m}$ represent the sets of complex and real matrices of size  $n \times m$, respectively. Bold-font variables denote matrices or vectors. $x \sim {\cal CN}(\mu,\sigma^2)$ means variable $x$ follows the circularly-symmetric complex Gaussian distribution with mean $\mu$ and covariance $\sigma^2$. $(\cdot)^T$ and $(\cdot)^H$ denote the transpose and Hermitian of a vector or a matrix, respectively. $\Re \{\cdot \}$ and $\Im \{\cdot \}$ refer to the real and the imaginary parts of a complex number.

\section{System Model and Problem Formulation}

Text is an important type of source data, which can be sensed from speaking and typing, environmental monitoring, etc. By training DL models with these text data at cloud/edge platform, the DL models based IoT devices have the capability to understand text data and generate semantic feature to be transmitted to the center to perform intelligent tasks, i.e.,  intelligent assistants, human emotion understanding, and environment humid and temperature adjustment based on human preference \cite{gil2016internet}.

\begin{figure*}[ht]
	\centering
	\includegraphics[width=165mm]{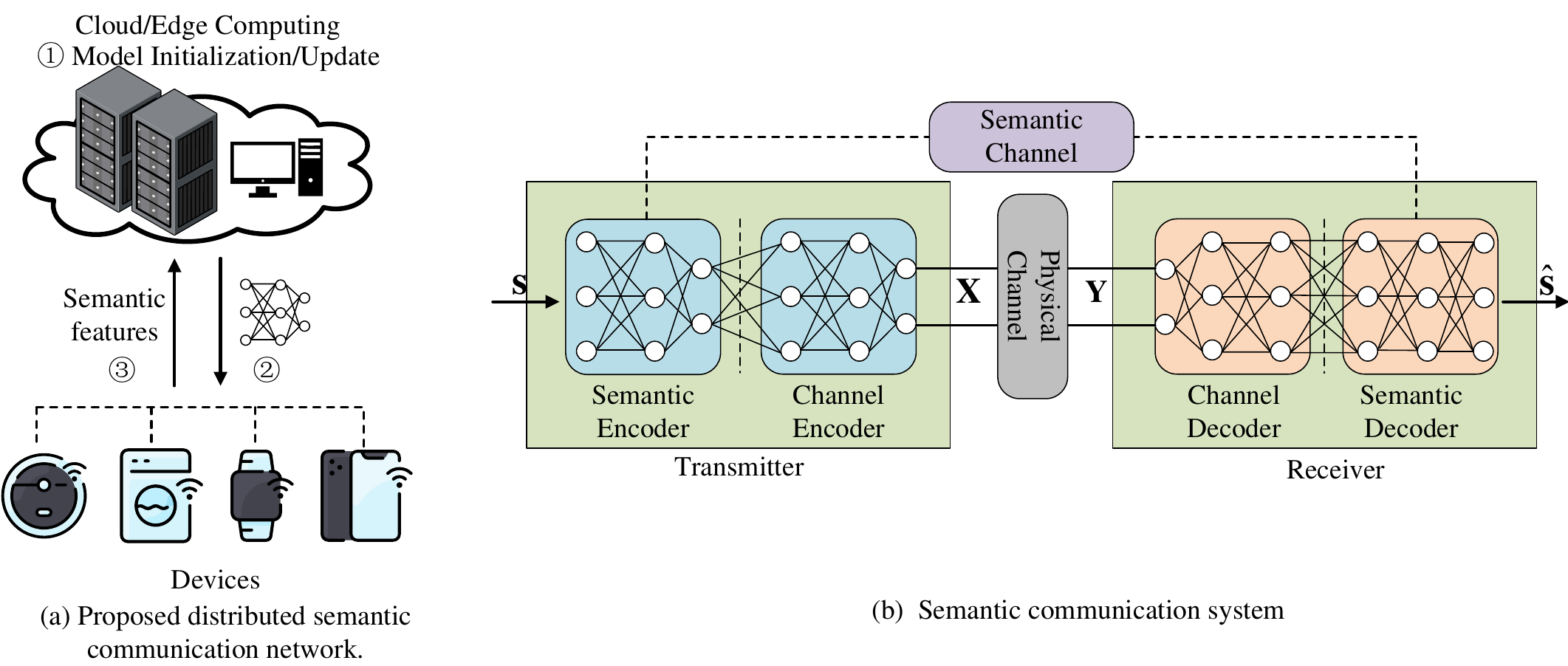}
	\caption{The framework of semantic communications for IoT networks.}
	\label{system model}
\end{figure*}

As shown in Fig. \ref{system model}(a), we focus on distributed semantic communications for IoT networks. The considered system is consisted of various IoT networks with two layers, the cloud/edge platform and distributed IoT devices. The cloud/edge platform is equipped with huge computation power and big memory, which can be used to train the DL model by the received semantic features. The semantic communication enabled IoT devices to perform intelligent tasks by understanding sensed texts, which are with limited memory and power but expected long lifetime, i.e., up to 10 years. Particularly, our considered distributed semantic communication system consists of the following three steps:
\begin{itemize}
    \item[1)] \textbf{Model Initialization/Update}: The cloud/edge platform first trains the semantic communication model by initial dataset. The trained model is updated in the subsequent iterations by the received semantic features from IoT devices.
    \item[2)] \textbf{Model Broadcasting}: The cloud/edge platform broadcasts the trained DL model to each IoT device.
    \item[3)] \textbf{Semantic Features Upload}: The IoT devices constantly capture the text data, which are encoded by the proposed semantic transmitter  shown in Fig. \ref{system model}(b). The extracted semantic features are then transmitted to the cloud/edge for model update and subsequent processing.
\end{itemize}
The aforementioned \textit{Questions 1-3} correspond to model initialization/update, semantic features uploading,  and model broadcasting, respectively. Different from the traditional information transmission, semantic features can be not only used for recovering the text at the semantic level accurately, but also exploited as the input of other modules, i.e., emotion classification, dialog system, and human-robot interaction, for training effect networks and perform various intelligent tasks directly.  The devices can also exchange semantic features, which has been previously discussed in our work in \cite{Xie2020deep}. We focus on the communication between cloud/edge platforms and local IoT devices to make the semantic communication model affordable. 

\subsection{Semantic Communication System}
The DeepSC shown in Fig. \ref{system model}(b) can be divided into three parts mainly, transmitter network, physical channel, and receiver network, where the transmitter network includes semantic encoder and channel encoder, and the receiver network consists of semantic decoder and channel decoder. 

We assume that the input of the DeepSC is a sentence, ${\mathbf{s}} = \left[ {{w_1},{w_2}, \cdots ,{w_N}} \right]$, where $w_n$ represents the $n$-th word in the sentence. The encoded symbol stream can be represented as
\begin{equation}\label{eq1}
    {\mathbf{X}} = {C_{{{ {\bm \alpha} }}}} \left( {{S_{{{\bm \beta} }}}\left( {\mathbf{s}} \right)} \right),
\end{equation}
where ${S_{ {\bm \beta}} }\left( \cdot \right)$ is the semantic encoder network with parameter set ${{\bm \beta}}$ and ${C_{\bm \alpha} }\left( \cdot \right)$  is the channel encoder with parameter set $\mathbf{ {\bm \alpha}}$. 

If $\bf X$ is sent through a wireless fading channel, the signal received at the receiver can be given by
\begin{equation}\label{eq2}
    {\bf{Y}} = {f_{\bf H} }({\bf{X}})={\bf HX} + {\bf N},
\end{equation}
where $\bf H$\footnote{Here, we have omitted discussion of complex channels. If the complex channel is $\bar {\bf H}$, then ${\bar {\bf H}} = \left[ {\Re \left( {\bf{H}} \right), - \Im \left( {\bf{H}} \right);\Im \left( {\bf{H}} \right),\Re \left( {\bf{H}} \right)} \right]$.} represents the channel gain between the transmitter and the receiver, and ${\bf N} \sim {\cal N}\left( {0,\sigma _n^2} \right)$ is additive white Gaussian noise (AWGN). 

The decoded signal can be represented as 
\begin{equation}
	{\mathbf{\hat s}} = {S^{-1}_{\bm \chi} } \left( {{C^{-1}_{\bm \delta} }\left( {\mathbf{Y}} \right)} \right),
\end{equation}
where $\bf \hat s$ is the recovered sentence, $C^{-1}_{\bm \delta} \left( \cdot \right)$ is the channel decoder with  parameter set ${\bm \delta}$ and $S^{-1}_{\bm \chi} \left( \cdot \right)$ is the semantic decoder network with  parameter set $\mathbf{ {\bm \chi}}$, the superscript -1 represents the decoding operation.

The whole semantic communication can be trained by the cross-entropy (CE) loss function, which is given by
\begin{equation}\label{loss function 1}
    \begin{aligned}
     {{\cal L}_{{\tt {CE}}}}({\bf{s}},{\bf{\hat s}}) = &\sum\limits_{i = 1} {\left( {q\left( {{w_i}} \right) - 1} \right)\log \left( {1 - p\left( {{w_i}} \right)} \right)} \\
     &- \sum\limits_{i = 1} {q\left( {{w_i}} \right)\log \left( {p\left( {{w_i}} \right)} \right)},
     \end{aligned}
\end{equation}
where $q(w_i)$ is the real probability that the  $i$-th word, $w_i$, appears in source sentence $\bf s$, and  $p({w_i})$ is the predicted probability that the  $i$-th word, $ w_i$, appears in  $\hat {\bf s}$. CE can measure the difference between the two distributions. Through minimizing the CE loss, the network can learn the word distribution, $q(w_i)$, in the source sentence, $\bf s$. Consequently, the syntax, phrase, and the meaning of words in the context can be learnt by DNNs.

\subsection{Problem Description}

Instead of bits, the input sentence, $\bf s$, in the DeepSC, will cause that the learned constellation is no longer limited to a few points anymore. After transmitting $\bf X$, the fading channel increases the difficulty of model training compared with the AWGN channel. Meanwhile, the huge number of parameters, ${\bm \alpha}, {\bm \beta}, {\bm \chi}, {\bm \delta}$, indicates the complexity of the whole model. These factors limit DeepSC for IoT networks and incur the aforementioned \textit{Questions 1-3}, including feasible constellation design, training for fading channel, and model compression.

\subsubsection{Training of fading channel}
In DL, the training process can be divided forward-propagation to predict the target and back-propagation to converge the neural network, as stated in the following.

\textbf{Forward-propagation}: From the received signal to recover semantic information, the estimation sentence is given by 
\begin{equation}\label{eq5}
    {\mathbf{\hat s}} = {S^{-1}_{\bm \chi} } \left( {{C^{-1}_{\bm \delta} }\left( {\mathbf{Y}} \right)} \right),
\end{equation}
\,\, \textbf{Back-propagation}: Taking semantic encoder as an example, the parameter vector at the $t_{th}$ iteration are is updated by
\begin{equation}
        {{\bf{\bm \beta }}(t)} = {{\bf{\bm \beta }}(t-1)} - \eta \frac{{\partial {\cal L}_{\tt CE}}}{{\partial {\bf{\bm \beta }}}},
\end{equation}
where  $\eta$ is the learning rate and $\frac{{\partial {\cal L}_{\tt CE}}}{{\partial {\bf{\bm \beta }}}}$ is the  gradient, computed by
\begin{equation}\label{gradients}
        \begin{aligned}
        \frac{{\partial {\cal L}_{\tt CE}}}{{\partial {\bf{\bm \beta }}}} &= \frac{{\partial {{\cal L}_{\tt CE}}}}{{\partial {\bf{\hat s}}}} \frac{{\partial {\bf{\hat s}}}}{{\partial {\bf{Y}}}}\frac{{\partial {\bf{Y}}}}{{\partial {\bf{X}}}}\frac{{\partial {\bf{X}}}}{{\partial {\bf{\bm \beta }}}}\\
         &= \frac{{\partial {{\cal L}_{\tt CE}}}}{{\partial {\bf{\hat s}}}} \frac{{\partial {\bf{\hat s}}}}{{\partial {\bf{Y}}}}{\bf{H}}\frac{{\partial {\bf{X}}}}{{\partial {\bf{\bm \beta }}}}.
         \end{aligned}
\end{equation}
    
In \eqref{gradients}, $\bf H$ will introduce stochasticity during weight updating. For an AWGN channel, $\bf H = I$  will not affect it. However, for fading channels, $\bf H$ is random, which may lead to that $\bm \beta$ fails to converge to the global optimum while the forward-propagation in \eqref{eq5} is unable to recover semantic information accurately based on the local optimum. Thus, it is critical to design the training process to mitigate the effects of $\bf H$, which also makes the DeepSC applicable for fading channels.

\subsubsection{Feasible constellation design}
Generally, the DL models run on floating-point operations (FLOPs), which means that the input, output, and weights are in a large range of $\pm 1.40129 \times {10^{ - 45}}$ to $\pm 3.40282 \times {10^{ + 38}}$ \cite{4610935}. Although DeepSC can learn the constellations from the source information and channel statistics,  the learned constellation points, such as cluster constellation \cite{zhu2019joint}, are disordered in the range of $\pm 1.40129 \times {10^{ - 45}}$ to $\pm 3.40282 \times {10^{ + 38}}$, which brings additional burden to the hardware of IoT devices, for instance, the high-resolution phase-shift and amplitude-shift pose high requirements on the circuit. Therefore, it is desired to form feasible constellations with only finite points for the current radio frequency (RF) systems. In other words, we have to design a smaller constellation for the DeepSC.

\subsubsection{Model communication}
The more parameters DeepSC has, the stronger the signal processing ability, which however increases computational complexity and model size and results in high power consumption. In the distributed DeepSC system, the trained DeepSC model deployed at local IoT devices is frequently updated to perform intelligent tasks better. The IoT application limits the bandwidth and cost of distributing the DeepSC model.  Furthermore, to extend the IoT network lifetime, especially the battery lifetime, most local devices are with finite storage and computation capability, which limits the size of DeepSC. Therefore, compressing DeepSC not only reduces the latency of model transmission between the cloud/edge platform and local devices but also makes it possible to run the DL model on local devices.

\section{Proposed Lite Distributed Semantic Communication System}
To address the identified challenges in Section II, we propose a lite distributed semantic communication system, named L-DeepSC. We analyze the effects of CSI in the model training under fading channels and design a CSI-aided training process to overcome the fading effects, which successfully deals with \textit{Question 1}. Besides, the weight pruning and quantization are investigated to address \textit{Question 2}. Finally, our finite-points constellation design solves \textit{Question 3}, effectively.

\subsection{Deep De-noise Network based CSI Refinement and Cancellation}
The most common method to reduce the effects of fading channels in wireless communication is to use the known channel properties of a communication link, CSI. Similarly, CSI can also reduce the channel impacts in training L-DeepSC.  Next, we will first analyze the role of CSI in L-DeepSC training.

In order to simplify the analysis, we assume the transmitter and the receiver are with one-layer dense with sigmoid activation, where transmitter has an additional untrainable embedding layer, and receiver also has an untrainable de-embedding layer. The IoT devices are with the trained transmitter model and the cloud/edge platform works as the receiver, as shown in the system model Fig. \ref{system model}.  The IoT devices and  cloud/edge platform are equipped with the same number of antennas. After the embedding layer, the source message, $\bf s$, is embedded into, $\bf S$. Then, encode $\bf S$ into
\begin{equation}
    {\bf{X}} = \sigma \left( {{{\bf{W}}_T}{\bf{S}} + {{\bf{b}}_T}} \right),
\end{equation}
where ${\bf{X}}$\footnote{Here, we have avoided discussion of complex signal. If the complex signal is $\bar {\bf X}$, then ${\bar {\bf X}} =  \left[ {\Re \left( {\bf{X}} \right), \Im \left( {\bf{X}} \right)} \right]. $} is the semantic features transmitted from the IoT devices to the cloud/edge platform. ${{\bf{W}}_T}$ and ${{\bf{b}}_T}$ are the trainable parameters to extract the features from source message $\bf s$, and $\sigma(\cdot)$ is the sigmoid activation function.

The received symbol at the cloud/edge platform is affected by channel $\bf H$ and AWGN as in \eqref{eq2}. From the received symbol, the cloud/edge platform recovers the embedding matrix by
\begin{equation}
    {\bf{\hat S}} = \sigma \left( {{{\bf{W}}_R}{\bf{Y}} + {{\bf{b}}_R}}  \right),
\end{equation}
where the estimated source message, $\hat{\bf s}$, can be obtained after de-embedding layer. ${{\bf{W}}_R}$ and ${{\bf{b}}_R}$ can learn to recover $\bf s$. The L-DeepSC can be optimized by the loss function in \eqref{loss function 1}. The fading channels not only contaminates the gradients in the back-propagation, but also restricts the representation power in the forward-propagation.

\textbf{Back-propagation}: It updates parameter ${{\bf{W}}_T}$ by its gradient
\begin{equation}\label{eq11}
    \frac{{\partial {\cal L}_{\tt CE}\left( {{\bf{\hat s}},{\bf{s}}} \right)}}{{\partial {{\bf{W}}_T}}} = {\left( {{{\bf{F}}_R}{{\bf{W}}_R}{\bf{H}}{{\bf{F}}_T}} \right)^T}{\nabla _{{\bf{\hat s}}}}{\cal L}_{\tt CE}\left( {{\bf{\hat s}},{\bf{s}}} \right){{\bf{s}}^T},
\end{equation}
where ${{\bf{F}}_R} \sim \text{diag}\left( {\sigma '\left( {{{\bf{W}}_R}{\bf{y}} + {{\bf{b}}_R}} \right)} \right)$ and ${{\bf{F}}_T} \sim \text{diag}\left( {\sigma '\left( {{{\bf{W}}_T}{\bf{s}} + {{\bf{b}}_T}} \right)} \right)$. In \eqref{eq11}, the $\bf H$ is untrainable and random, therefore it will cause perturbation for the weight updating, i.e., the weight updating with higher variance. If the transmitter consists of very deep neural networks, the perturbation will affect the back-propagation of the whole transmitter network, where the perturbation will propagate to the whole transmitter network by the chain rule.

\textbf{Forward-propagation}: With the received signal ${\bf W}_R$, the source messages can be recovered by
\begin{equation}\label{eq12}
    \begin{aligned}
{{\hat {\bf S}}} &= \sigma \left( {{\bf{W}}_R}{\bf{Y}} + {{\bf{b}}_R} \right)\\
 &= \sigma \left( {{\bf{W}}_R}{\bf{HX}} + {{\bf{W}}_R}{\bf{N}} + {{\bf{b}}_R} \right).
\end{aligned}
\end{equation}
In \eqref{eq12},  ${\bf W}_R$ has to learn how to deal with the channel effects and decode at the same time, which increases training burden and reduces network expression capability. Meanwhile, the errors caused by channel effects also propagate to the subsequent layers for the L-DeepSC receiver with multiple layers.

The impacts of channel can be mitigated by exploiting CSI at the cloud/edge. If channel $\bf H$ is known, then the received symbol can be processed by
\begin{equation}\label{eq13}
    \begin{aligned}
    {\bf{\tilde Y}} &= {\left( {{{\bf{H}}^H}{\bf{H}}} \right)^{ - 1}}{{\bf{H}}^H}{\bf{Y}}
     &= {\bf{X}} + {\bf{\tilde N}},
     \end{aligned}
\end{equation}
where ${\bf{\tilde N}} = {\left( {{{\bf{H}}^H}{\bf{H}}} \right)^{ - 1}}{{\bf{H}}^H}{\bf{N}}$. In \eqref{eq13},  the channel effect is transferred from multiplicative noise to additive noise, ${\bf{\tilde N}}$, which provides the possibility of stable back-propagation as well as the stronger capability of network representation. With \eqref{eq13}, back-propagation and forward-propagation can be performed by setting $\bf H = I$ in \eqref{eq11} and \eqref{eq12}, respectively. Therefore, the channel effects can be completely removed.

The above discussion shows the importance of CSI in model training. However, CSI can be only estimated generally, i.e., least-squared (LS), linear minimum mean-squared error (LMMSE), or minimum mean-squared error (MMSE) estimators. Due to exploiting prior channel statistics, LMMSE and MMSE estimators usually perform better than the LS estimators. Thus, LMMSE and MMSE estimators are sensitive to the accuracy of channel statistic while the LS estimator requires no prior channel information. Meanwhile, DL techniques can also be used to improve the performance of channel estimation \cite{Thakkar9074414, BaleviDA20}.

\begin{figure}[!t]
	\centering
	\includegraphics[width=90mm]{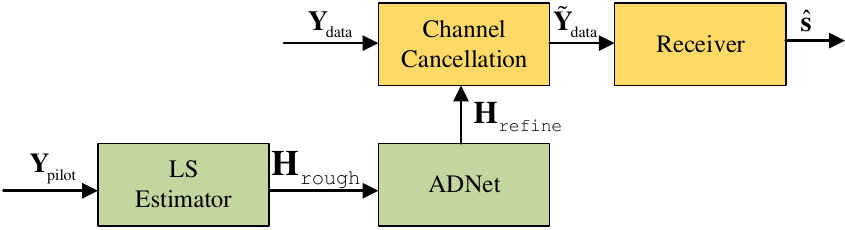}
	\caption{The proposed CSI refinement and cancellation based on de-noise neural networks.}
	\label{ADNet}
\end{figure}

For simplicity, we initially use the LS estimator. Then, we adopt the deep de-noise network to increase the resolution of the LS estimator as in \cite{zhang2017beyond} shown in Fig. \ref{ADNet}. Particularly, the rough CSI estimated by the LS estimator with few pilots first denoted by 
\begin{equation}\label{eq132}
{\bf{ H}}_{\tt rough} = {{\bf Y}_p {\bf X}}_p^H = {\bf{H}} + {\bf{NX}}_p^H,
\end{equation}
where ${{\bf{Y}}_p} = {\bf{H}}{{\bf{X}}_p}{\bf{ + N}}$, ${{\bf{Y}}_p}$ is the received pilot signal, ${{\bf{X}}_p}$ is the transmitted pilot signals. Then,  \eqref{eq132} can be represented as
\begin{equation}\label{eq16}
    {\bf{ H}}_{\tt rough} = {\bf{H}} + {\bf \widehat {N}},
\end{equation}
where ${\bf \widehat{N}} = {\bf{NX}}_p^H$.

From \eqref{eq16}, ${\bf{ H}}_{\tt rough}$ consists of exact $\bf H$ and the noise, $\bf \widehat N$.  De-noise neural networks are used to recover $\bf H$ more accurately from ${\bf H}_{\tt rough}$ by considering $\bf H$ and ${\bf{ H}}_{\tt rough}$ as the original picture and noisy picture, respectively. Here, we exploit attention-guided denoising convolutional neural network (ADNet) \cite{tian2020attention} to refine CSI. ADNet includes four blocks, a sparse block, a feature enhancement block, an attention block, and a reconstruction block. After the input image, the sparse block is used to extract useful features from the given noisy image. Attention block can extract the noise information hidden in the complex background and is integrated into the feature enhancement block to reduce the complexity. Finally, the de-noised image is reconstructed by the reconstruction block.

The refined CSI, ${{\bf{H}}_{\tt refine}}$ denoted by
\begin{equation}\label{eq17}
    {{\bf{H}}_{\tt refine}} = \text{ADNet}\left( { {\bf{ H}}_{\tt rough}} \right).
\end{equation}
In \eqref{eq17}, the ADNet$(\cdot)$ is trained the the loss function, ${\cal L}\left( {{{\bf{H}}_{\tt refine}},{\bf{H}}} \right) = \frac{1}{2}\left\| {{{\bf{H}}_{\tt refine}} - {\bf{H}}} \right\|_F^2$.  Since the performance of the LS estimator is similar to that of LMMSE and MMSE estimators in the high SNR region, we pay more attention to the low SNR region when training ADNet. With proper training, ADNet can mitigate the impacts from noise but without any prior channel information, especially in the low SNR region. Such a design provides a good solution for \textit{Question 1}.

\subsection{Model Compression}
\begin{figure}[!t]
	\centering
	\includegraphics[width=90mm]{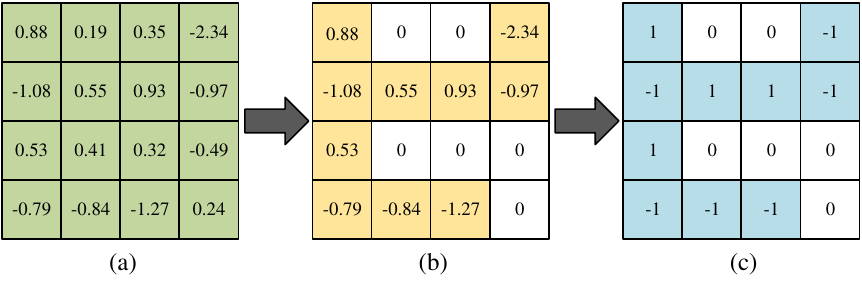}
	\caption{Flowchart of the proposed joint pruning-quantization, (a) the original weights matrix; (b) the weights after pruning, where the example pruning function is $x=0$ for $x<0.5$; (c) the weights after quantization, where the example quantization function is $x = \text{sign}(x)$.}
	\label{Flowchart}
\end{figure}

Through applying CSI into model training,  the cloud/edge platform can extract the semantic features from L-DeepSC. However, the size and complexity of the trained L-DeepSC model are still very large, which causes high latency for the cloud/edge platform to broadcast updated L-DeepSC. Note that both weights pruning and quantization can reduce the model size and complexity, therefore, we compress the DeepSC model by a joint pruning-quantization scheme to make it affordable for IoT devices. As shown in Fig. \ref{Flowchart}, the original weights are first pruned at a high-precision level by identifying and removing the unnecessary weights, which makes the network sparse. Quantization is then used to convert the trained L-DeepSC model into a low-precision level. The proposed network sparsification and quantization can address \textit{Question 3} and are introduced in detail in the following.

\subsubsection{Network Sparsification}
A proper criterion to disable neural connections is important. Obviously, the connections with small weight values can be pruned. Therefore, the pruning issue here turns into setting a proper pruning threshold. 

As shown in Fig. \ref{system model}(b), the DeepSC consists with neural networks, $\bm \alpha, \bm \beta, \bm \chi, \bm \delta$, where each includes multiple layers. As the DeepSC mainly consists of dense layers, we choose unstructured pruning method in this paper, where the computation workload of sparse model can be reduced by the sparsity algorithm and FPGA design \cite{dorrance2014scalable, zhuo2005sparse}, i.e., sparse matrix-vector multiplication. Assume there are total $N$ layers in the pre-trained DeepSC model with ${\bf W}^{(n)}_{i,j}$ being the weight of connection between the $i_{th}$ neuron of the $(n+1)_{th}$ layer and $j_{th}$ neuron of $n_{th}$ layer. With a pruning threshold $w_{\tt thre}$, the model weights can be pruned by
\begin{equation}\label{prune}
    {\bf{W}}_{i,j}^{\left( n \right)} = \left\{ \begin{array}{l}
            {\bf{W}}_{i,j}^{\left( n \right)},\, {\rm{ if }} \, \left| {{\bf{W}}_{i,j}^{\left( n \right)}} \right| > {w_{\tt thre}},\\
            0,\qquad {\rm{ otherwise}},
\end{array} \right.
\end{equation}
We determine the pruning threshold by
\begin{equation}\label{thre}
    w_{\tt thre} = {\bf s}_{M\times \gamma},
\end{equation}
where ${\bf{s}} = {\rm{sort}}\left(  \left[ {{{\bf{W}}^{\left( 1 \right)}},{{\bf{W}}^{\left( 2 \right)}}, \cdots ,{{\bf{W}}^{\left( N \right)}}} \right] \right)$, is the sorted weights value from least important one to the most important one, $M$ is the total number of connections, and $\gamma$, the sparsity ratio between 0 and 1, indicates the proportion of zero values in weights.  The weight pruning can be divided into two steps, weight pruning to disable some neuron connections and fine-tine to recover the accuracy, as shown in Algorithm \ref{weights pruning}.

\begin{algorithm}[!t]
	\caption{Network Sparsification.} 
	\label{weights pruning} 
	\text{\textbf{Input}: The pre-trained weights $\bf W$, the sparse ratio $\gamma$.}
	
	\text{\textbf{Output}: The pruned weights ${\bf W}_{\tt pruned}$.}
	
	\begin{algorithmic}[1] %这个1 表示每一行都显示数字
	\STATE  Count the the total number of connections, $M$.
	\STATE  Sort the whole connections from small to large, $\bf s$.
	\STATE  Obtain the threshold by \eqref{thre} with $M$ and $\gamma$, $w_{\tt thre}$.
	\FOR{$n=1$ to $N$}
    \STATE  Prune the connections by \eqref{prune},  ${\bf W}^{(n)}_{\tt pruned}$.
    \ENDFOR
	\STATE Fine-tune the pruned model by loss function \eqref{loss function 1}.
	\end{algorithmic}
\end{algorithm}

\subsubsection{Network Quantization}
The quantization includes weight quantization and activation quantization. The weights, ${\bf W}^{(n)}_{i,j}$, from a trained model, can be converted from 32-bit float point to $m$-bits integer through applying the quantization function by
\begin{equation}\label{quantization equation}
    {\bf{\tilde W}}_{i,j}^{(n)} = \text{round}\left( {{q_w}\left( {{\bf{W}}_{i,j}^{(n)}  - \min \left({{\bf{W}}^{(n)} } \right) } \right)} \right),
\end{equation}
where $q_w$ is the scale-factor to map the dynamic range of float points to an $m$-bits integer, which is given by
\begin{equation}\label{eq21}
    {q_w} = \frac{{{2^m} - 1}}{{\max \left( {{{\bf{W}}^{(n)} }} \right) - \min \left( {{{\bf{W}}^{(n)} }} \right)}}.
\end{equation}

\begin{algorithm}[!t]
	\caption{Network Quantization.} 
	\label{weights quantization}
	\text{\textbf{Input}: The pre-trained weights $\bf W$, the quantization level $m$, }\\
	 \text{\qquad \quad  the correlation coefficient $c$, and the calibration data $\cal K$.}\\
	\text{\textbf{Output}: The pre-trained weights ${\bf W}_{\tt quantized}$ and the range of  }\\
	\text{\qquad \quad activation  $x_{\min}$ and $x_{\max}$.}
	
	\begin{algorithmic}[1] 
	\STATE \text{\textbf{Phase 1:} Weights Quantization.}
	\FOR{$n=1$ to $N$}
	\STATE  Compute the range of weights, ${\max \left( {{{\bf{W}}^{(n)}}} \right)}$ and ${\min \left( {{{\bf{W}}^{(n)}}} \right)}$.
    \STATE  Quantize the weights by \eqref{quantization equation}, ${{{\bf{\tilde W}}}^{(n)}}$.
    \ENDFOR
    \vspace{0.55em}
    \STATE \text{\textbf{Phase 2:} Activations Quantization.}
    \FOR{$t=1$ to $\cal K$}
    \FOR{$n=1$ to $N$}
	\STATE  Update the dynamic range of activation by \eqref{ema} and \eqref{ema2}, $x^{(n)}_{\min}(t)$ and $x^{(n)}_{\max}(t)$.
    \ENDFOR
    \ENDFOR
    \STATE Quantize the activations by \eqref{activations quantization}.
	\STATE  Fine-tune the quantized model by STE and loss function \eqref{loss function 1}.
	\end{algorithmic}
\end{algorithm}

For activation quantization, the results of matrix multiplication are stored in accumulators. Due to the limited dynamic range of integer formats, it is possible that the accumulator overflows quickly if the bit-width for the weights and activation is the same. Therefore, accumulators are usually implemented with higher bit-widths, for example, INT32 += INT8$\times$ INT8. Besides, the range of activations is dynamic and dependent on the input data. Therefore, the output of activations has to re-quantize into $m$-bits integer for the subsequent calculation. Unlike weights that are constant, the output of activations usually includes elements that are statistical outliers, which expand the actual dynamic range. For example, even if 99\% of the data is distributed between -100 and 100, an outlier, 10,000, will extend the dynamic range into from -100 to 10,000, which significantly reduces the mapping resolution. In order to reduce the influence from the outliers, an exponential moving average (EMA) is used by
\begin{align}
        {x^{(n)}_{\min }}(t+1) &= \left( {1 - c} \right){x^{(n)}_{\min }}(t) + c\min \left( {{{\bf X}^{(n)}}\left( t \right)} \right),  \label{ema}
 \end{align}
and
\begin{align}
        {x^{(n)}_{\max }}(t+1) &= \left( {1 - c} \right){x^{(n)}_{\max }}(t) + c\max \left( {{{\bf X}^{(n)}}\left( t \right)} \right),\label{ema2}
\end{align}
where ${x^{(n)}_{\min }}(t+1)$ and ${x^{(n)}_{\max }}(t+1)$ are used for the range of activation quantization, and  ${x^{(n)}_{\min }}(1)=\min \left( {{{\bf X}^{(n)}}\left( 1 \right)} \right)$, ${x^{(n)}_{\max }}(1)=\max \left( {{{\bf X}^{(n)}}\left( 1 \right)} \right)$, ${\bf{X}}^{(n)}(t)$ is the output of activations at $n_{th}$ layer with $t_{th}$ batch data, $c \in [0,1)$ represents the correlation between the current $x^{(n)}_{\min}$/$x^{(n)}_{\max}$ with its past value. The effects from outliers can be mitigated by the past normal values. After $t+1$ epochs, the $x^{(n)}_{\min}$ and $x^{(n)}_{\max}$ are fixed based on $x^{(n)}_{\min}(t+1)$ and $x^{(n)}_{\max}(t+1)$. Then, the output of the activations can be quantized by
\begin{equation}\label{activations quantization}
    {\bf{\tilde X}}^{(n)} = \text{clamp}\left( {\text{round}\left( {{q_x}\left( {{\bf{X}}^{(n)} - {x^{(n)}_{\min }}} \right)} \right); - M,M} \right),
\end{equation}
where ${q_a} = ({2^m} - 1)/({x^{(n)}_{\max }} - {x^{(n)}_{\min }})$ is the scale-factor and $\text{clamp}\left( \cdot \right)$ is used to eliminate the quantized outliers, which is given by
\begin{equation}
    \text{clamp}\left( {{\bf{X}}^{(n)}; - T,T} \right) = \min \left( {\max \left( {{\bf{X}}^{(n)}, - T} \right),T} \right),
\end{equation}
where $T= 2^m-1$, which is the border of the $m$-bits integer format. 

As shown in Algorithm \ref{weights quantization}, the network quantization includes two phases: i) weight quantization; ii) activations quantization. In phase 1, the weights of each layer can be quantized by \eqref{quantization equation} directly. In phase 2, the calibration process is applied by running a few calibration batches in order to get the activations statistics. In each batch, $x^{(n)}_{min}(t)$ and $x^{(n)}_{max}(t)$ will be updated based on the activations statistics from the previous batches. These quantization processes might lead to slight accuracy degradation. The quantization-aware training (QAT)  is required to re-train for minimizing the loss of accuracy.  Since the rounding operation is not derivable, a straight-through estimator (STE) is used to estimate the gradient of quantized weights in the back-propagation \cite{bengio2013estimating}.

\subsection{Constellation Design with Fewer Quantization Bits}
The cloud/edge platform can further reduce the size of L-DeepSC with model compression after the model is trained, which not only reduces the latency significantly for broadcasting the updated DeepSC to IoT devices, but also changes DeepSC to  L-DeepSC with low complexity. However, high-resolution waveform poses high requirements cost-sensitive IoT devices.  In other words, the cost-sensitive IoT devices are usually capacity-limited and cannot afford a large number of constellation points which are with phase and amplitude close to each other.

Different from bits, the source message, $\bf s$, is more complicated and the learned constellation will not be limited to a few points, which brings additional burden on hardware. Besides, the DL models generally run in FP32, which also expands the range of constellation. Thus, we aim to reduce the size of learned constellation without degrading performance, where the output of $\bf X$ is the learned constellation while $\bf X$ is also the output of activation of last layer at the local IoT devices. Inspired from the network quantization, we convert the learned high-resolution constellation into low-resolution one with few points. Thus, we use two-stage quantization to narrow the range of constellations, which is represented by
\begin{equation}\label{js}
    {{\bf{X}}_{{\tt {dequantize}}}} = \frac{{{{{\bf{ X}}}_{\tt quantize}}}}{{{q_x}}} + {x_{\min }},
\end{equation}
where ${\bf X}_{\tt quantize}$ is the quantized $\bf X$ from \eqref{activations quantization}, ${{{q_x}}}$ is the scale-factor and $x_{\min}$ is the obtained by \eqref{ema} and ${\bf X}_{\tt dequantize}$ is the dequantized $\bf X$.

First, we quantize the $\bf X$ into $m$-bits integer so that the range of $\bf X$ is narrowed to the size of $2^m$. For example, when $m= 8$, the size of the constellation is reduced to 256. Then, $\bf X_{\tt quantize}$ is dequantize to restore $\bf X$. Such an ${\bf X}_{\tt dequantize}$ has a similar distribution as $\bf X$ but is with fewer constellation points, which is helpful to lower the hardware cost at transmitter and preserves the performance as much as possible and therefore provides the solution for \textit{Question 2}.

In summary, by exploiting the solutions for the aforementioned \textit{Questions}, we develop a lite distributed semantic communication system, named L-DeepSC, which could reduce the latency for model exchange under limited bandwidth,  run the models at IoT devices with low power consumption, and deal with the distortion from fading channels when uploading semantic features. As a result, the proposed L-DeepSC becomes a good candidate for the IoT networks.

\section{Numerical Results}
In this section, we compare the proposed L-DeepSC with traditional methods under different fading channels, including Rayleigh and Rician fading channels. The weights pruning and quantization are also verified under fading channels. For the Rayleigh fading channel, the channel coefficient follows ${\cal CN}(0,1)$; for the Rician fading channel. it follows ${\cal CN}(\mu,\sigma^2)$ with $\mu = \sqrt{k/(k+1)}$ and $\sigma = \sqrt{1/(k+1)}$. where $k$ is the Rician coefficient and we use $k=2$ in our simulation.

The transmitter of L-DeepSC is the same as that of DeepSC in \cite{Xie2020deep}.  The parameters for the decoding network at the receiver are shown in Table \ref{table 2} for the fading channels, where the sum of the outputs of Dense 3 and Dense 5 is the input of the LayerNorm layer. The Transformer encoder and decoder are semantic encoder and decoder \cite{Xie2020deep}, respectively, which enables the systems to understand text and extract semantic information. We also prune the whole network since we consider the communications between cloud/edge platform and each IoT devices as well as the communications between IoT devices.

\begin{table}[htbp]
\footnotesize
\caption{The setting of L-DeepSC transceiver.}
\label{table 2}
\centering
\begin{tabular}{ |c| c |c |c |} 
\hline
& Layer Name&  Units &  Activation \\
\hline
\multirow{4}{4.5em}{\centering \\ Transmitter} & Embedding layer  & 128 & None \\
\cline{2-4}
& 4$\times$Transformer Encoder & 128 (8 heads) & None \\
\cline{2-4}
& Dense 1  & 256 & Relu \\
\cline{2-4}
& Dense 2 & 16 & None \\
\hline
\multirow{6}{4.5em}{\centering Receiver} & Dense 3 & 128 & Relu \\
\cline{2-4}
& Dense 4 & 512 & Relu \\
\cline{2-4}
& Dense 5 & 128 & None \\
\cline{2-4}
& LayerNorm & None & None \\
\cline{2-4}
& 4$\times$Transformer Decoder & 128 (8 heads) & None \\
\cline{2-4}
& Prediction Layer & Dictionary Size & Softmax \\
\hline
\end{tabular}
\end{table}

The output features are with 8 symbols per word. We initialize the learnable embedding matrix from ${\cal N}(0,1)$ with shape (vocab size, embedding-dim). The embedding dim is set to 128 in our program and  the vocab size depends on the training dataset. The batch size is 64, learning rate is ${128^{ - 0.5}}\min \left( {step^{ - 0.5},step \times {{4000}^{ - 1.5}}} \right)$, where \textit{step} is the counting number in the back-propagation. This corresponds to increasing the learning rate linearly for the first 4000 training steps and decreasing it thereafter proportionally to the inverse square root of the step number. We also adopt the $L_2$ regularization and the Adam optimizer with $\beta_1= 0.9$, $\beta_2 = 0.98$, and  $\varepsilon = 10^{-8}$.

The adopted dataset is the proceedings of the European Parliament \cite{koehn2005europarl}, which consists of around 2.0 million sentences and 53 million words. The dataset is pre-processed into lengths of sentences with 4 to 30 words and is split into training data and testing data with 0.1 ratio. The benchmark approach is based on separate source coding and channel coding technologies, which adopt variable-length coding (Huffman coding) for source coding, where we build the Huffman codes by counting the frequency of letters and punctuation so that the look-up table is not large. Turbo coding and Reed-Solomon (RS) coding \cite{reed1960polynomial} for channel coding, where turbo decoding method is log-MAP algorithm with 5 iterations, and quadrature amplitude modulation (QAM). The bilingual evaluation understudy (BLEU) score is used to measure the performance \cite{papineni2002bleu}.

\begin{figure*}[!t]
    \centering
    \subfigure[Full-resolution Constellation]{
    \includegraphics[width=80mm]{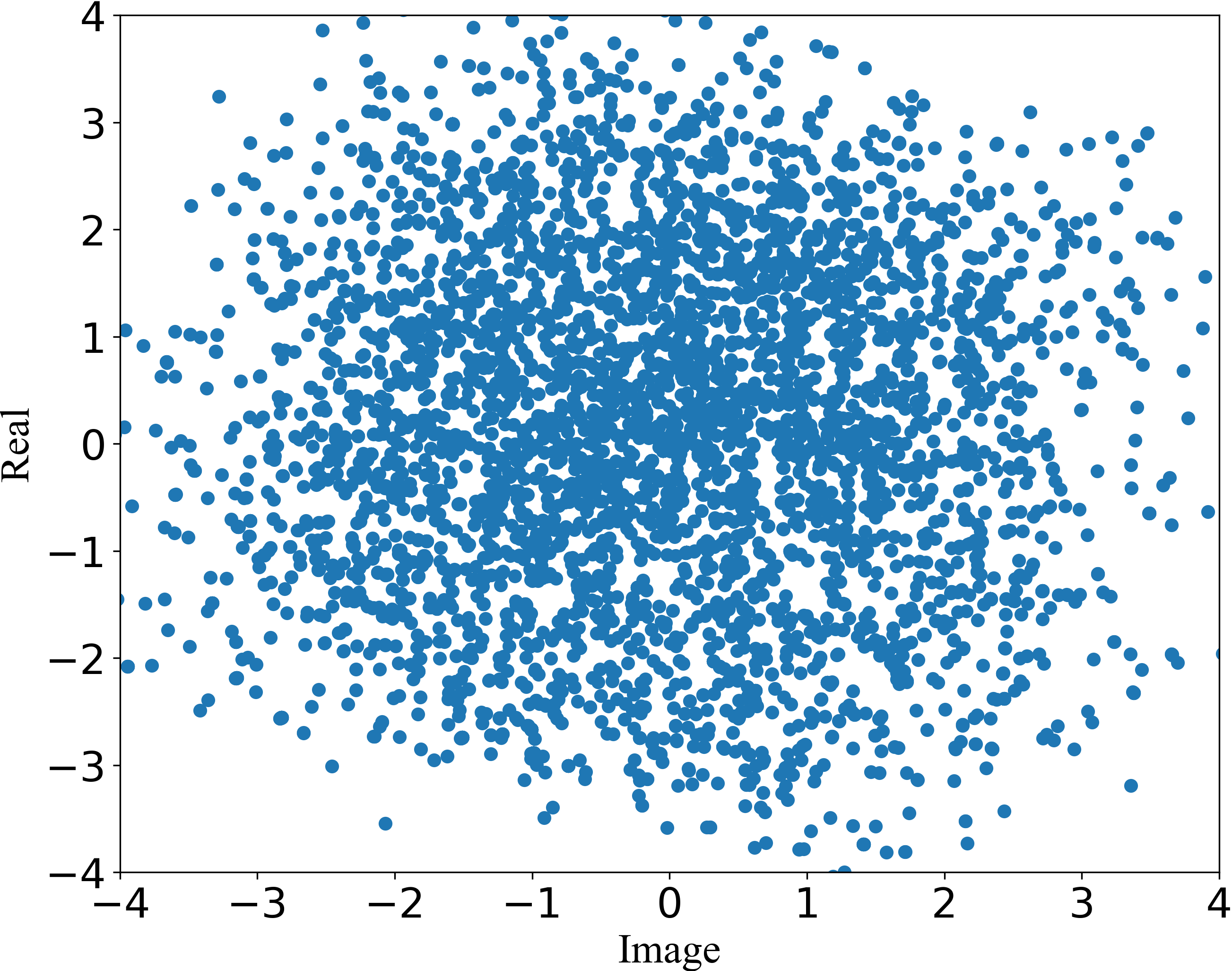}
    \label{constellations-a}
    }
    \subfigure[4-bits constellation]{
    \includegraphics[width=80mm]{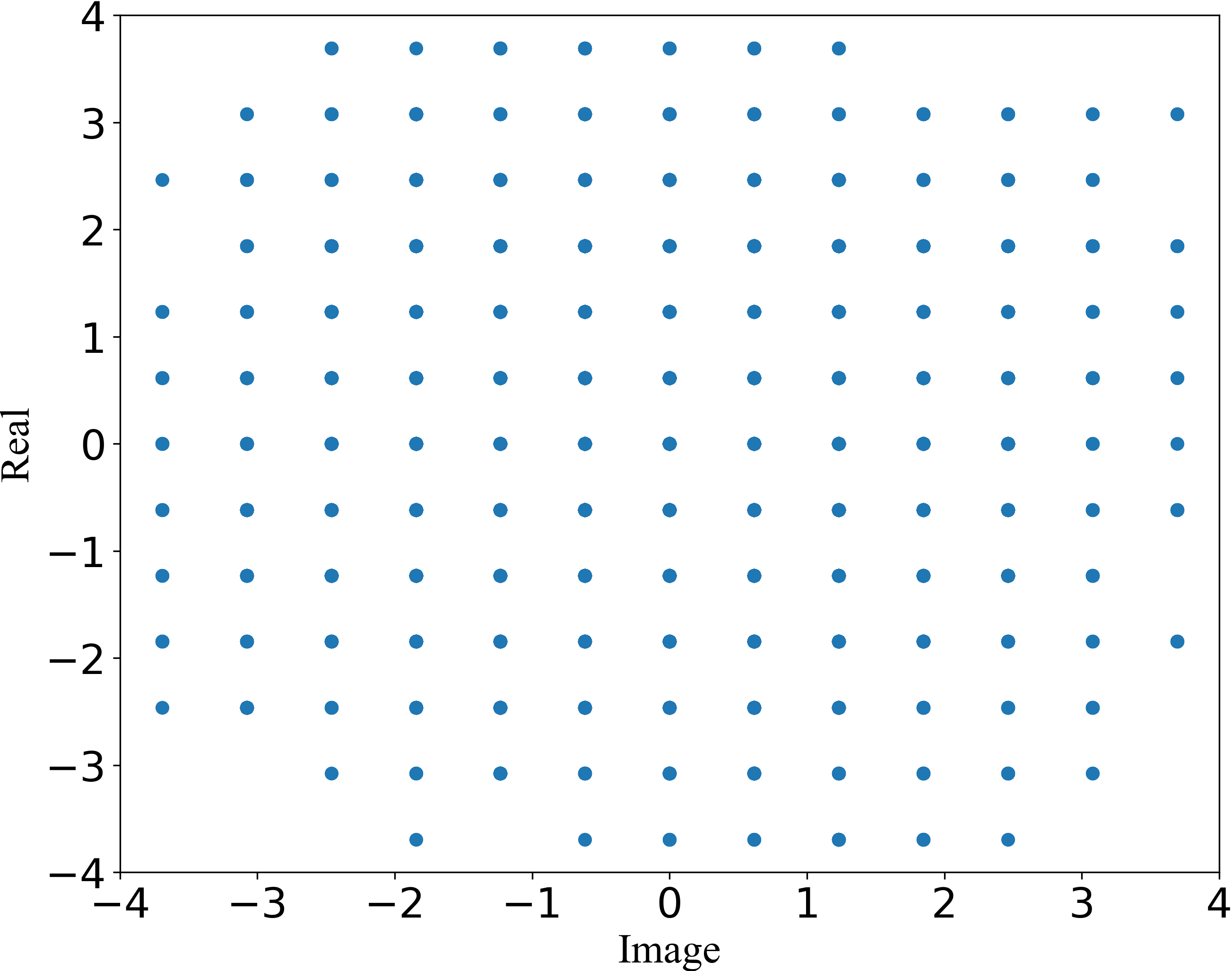}
    \label{constellations-b}}
    
    \caption{The comparison between the full-resolution constellation and 4-bits constellation.}
    \label{constellations}
\end{figure*}

% radio frequency
\subsection{Constellation Design}
Fig. \ref{constellations} compares the full-resolution constellation and the 4-bits constellation. The full-resolution constellation points in Fig. \ref{constellations-a} contain more information due to the higher resolution, but require complicated hardware, which is almost impossible to design. Through mapping the full-resolution constellation into a finite space, the 4-bits constellation points in Fig. \ref{constellations-b} become simplified, which makes it possible to implement in the existing RF system. Note that the 4-bits constellation keeps a similar distribution with the full-resolution constellation. For example, there exist certain blank regions at the edge of the constellation in Fig. \ref{constellations-a}, while the 4-bits constellation shows a similar trend in Fig. \ref{constellations-b}. Such similar distribution prevents sharp performance degradation  when the resolution of constellation decreases significantly. 

\begin{figure}[!t]
	\centering
	\includegraphics[width=85mm, height = 65mm]{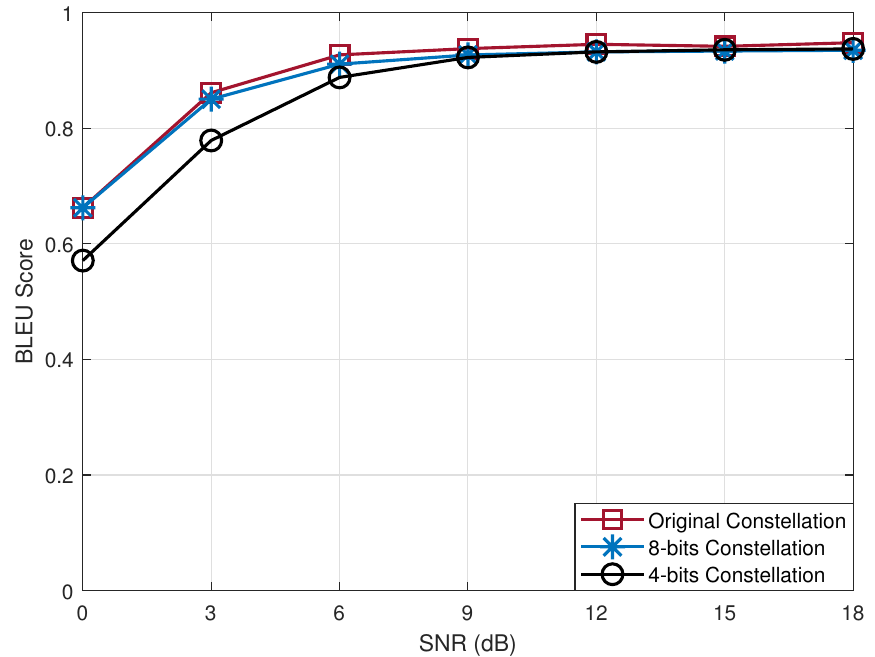}
	\caption{The BLEU scores of different constellation sizes versus SNR under AWGN.}
	\label{constellation performance}
\end{figure}

Fig. \ref{constellation performance} shows the BLEU scores versus SNR for different constellation sizes under AWGN, including 4-bits constellation, 8-bits constellation, and full-resolution constellation. All of them could achieve very similar performance when $\text{SNR}>9\,\text{dB}$, which demonstrates the constellation design is effective and cause no significant performance degradation. Full resolution and 8-bits constellations perform slightly better than 4-bits constellation when SNR is low. This is because some weights information used for denoising is lost when the resolution of the constellation is small.

\begin{figure}[!t]
	\centering
	\includegraphics[width=85mm, height = 65mm]{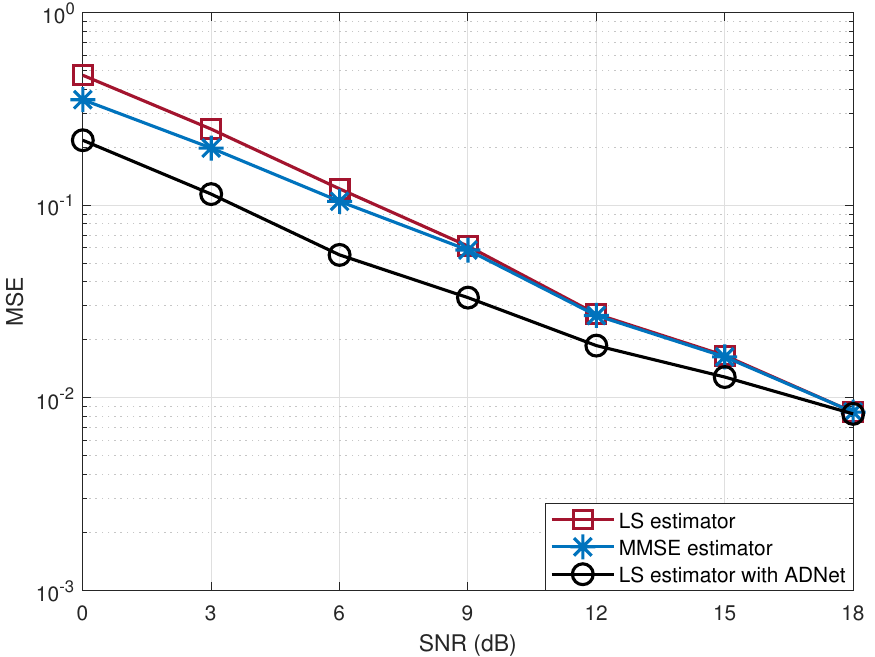}
	\caption{The MSE for MMSE estimator, LS estimator, and the proposed ADNet based LS estimator.}
	\label{MSE}
\end{figure}

\subsection{Performance over Fading Channels}
Fig. \ref{MSE} compares the channel estimation MSEs of LS, MMSE, and ADNet-aided LS estimator  versus SNR under the Rayleigh fading channels. Note that MMSE equals to LMMSE for the AWGN channels. The MMSE and LS estimators have similar accuracy in the high SNR region, thus the range of training SNRs for the ADNet is set from 0 dB to 10 dB to improve the performance of the LS estimator in the low SNR region. As a result, the MSE of ADNet based LS estimator is significantly lower than that of LS and MMSE estimators when SNR is low. With increasing SNR, the MSE of ADNet based LS estimator approaches to that of the LS and MMSE estimators. Therefore, the ADNet based LS estimator can be substituted by the LS estimator to reduce the complexity in the high SNR region.

\begin{figure}[!t]
	\centering
	\includegraphics[width=85mm, height = 65mm]{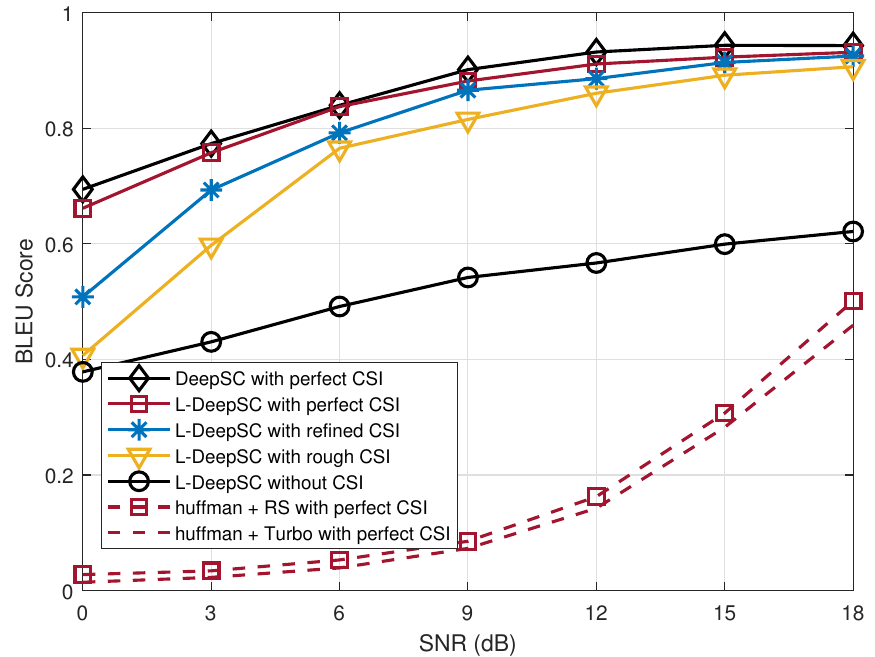}
	\caption{The BLEU scores versus SNR under Ricain fading channels, with perfect CSI, rough CSI, refined CSI, and no CSI.}
	\label{Rician}
\end{figure}

\begin{figure}[!t]
	\centering
	\includegraphics[width=85mm, height = 65mm]{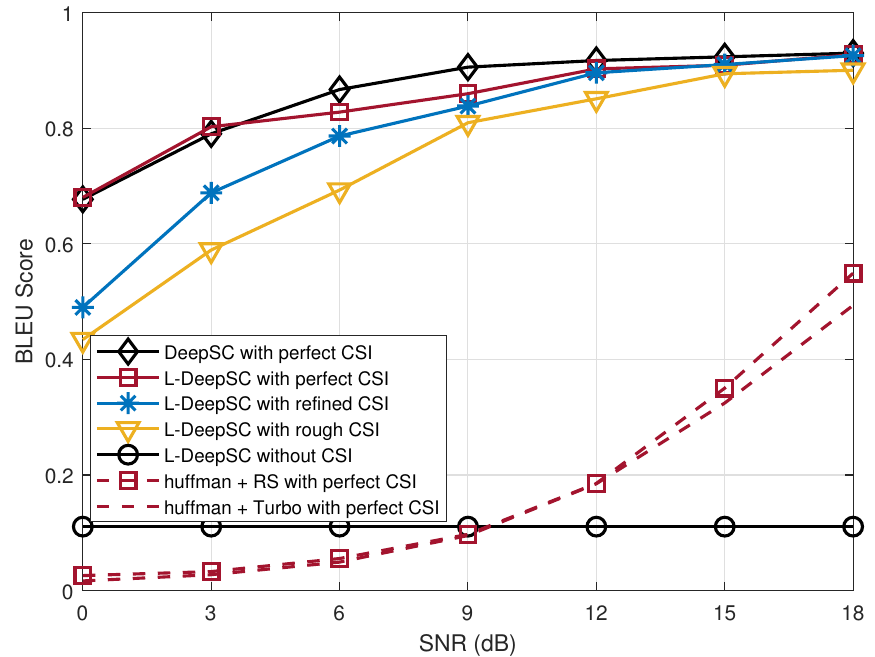}
	\caption{The BLEU scores versus SNR under Rayleigh fading channels, with perfect CSI, rough CSI, refined CSI, and no CSI.}
	\label{rayleigh}
\end{figure}

Fig. \ref{Rician} and Fig. \ref{rayleigh} illustrate the relationship between BLEU score and SNR with the 4-bits constellation over the Rician and the Rayleigh fading channels, respectively, where DeepSC is trained with perfect CSI and the L-DeepSC is trained with perfect CSI, rough CSI by \eqref{eq16}, refined CSI by \eqref{eq17} and without CSI, respectively. The traditional approaches are Huffman coding with (5,7) RS and with turbo coding (rate 1/2), both with 64-QAM. We observe that all DL-enabled approaches are more competitive under the fading channels. RS coding is better than turbo coding in terms of BLEU score. This is because RS coding is linear block coding with long block-length, which can correct long bit sequences, however, turbo coding is convolution coding with short block-length, where the coded bits only are related with previous $m$ bits, i.e., $m=3$, so that the adjacent words result in higher error rate. The performance of L-DeepSC is very close to that of DeepSC in terms of BLEU score, but requires much less bandwidth for communications. The system trained without CSI performs worse than those trained with CSI, especially under the Rayleigh fading channels, which also confirms the analysis of \eqref{eq11} and \eqref{eq12}. Without CSI, the performance difference  between the Rayleigh channels and the Rician channels is caused by the line-of-sight (LOS), which can help the systems recognize the semantic information during training. Besides, with the aid of CSI, the effects of the fading channels are mitigated significantly, as we have analyzed before. When SNR is low, the system with perfect CSI or refined CSI outperforms that with rough CSI. As SNR increases, all these systems, L-DeepSC with perfect CSI, refined CSI, and rough CSI, converge to similar performance gradually.

\subsection{Model Compression}
In this experiment, we investigate the performance of network slimmer, including network sparsification, network quantization, and the combination of both. The pre-trained model used for pruning and quantization is trained with 4-bits constellation under the Rician fading channels.

\begin{figure}[!t]
	\centering
	\includegraphics[width=85mm, height = 65mm]{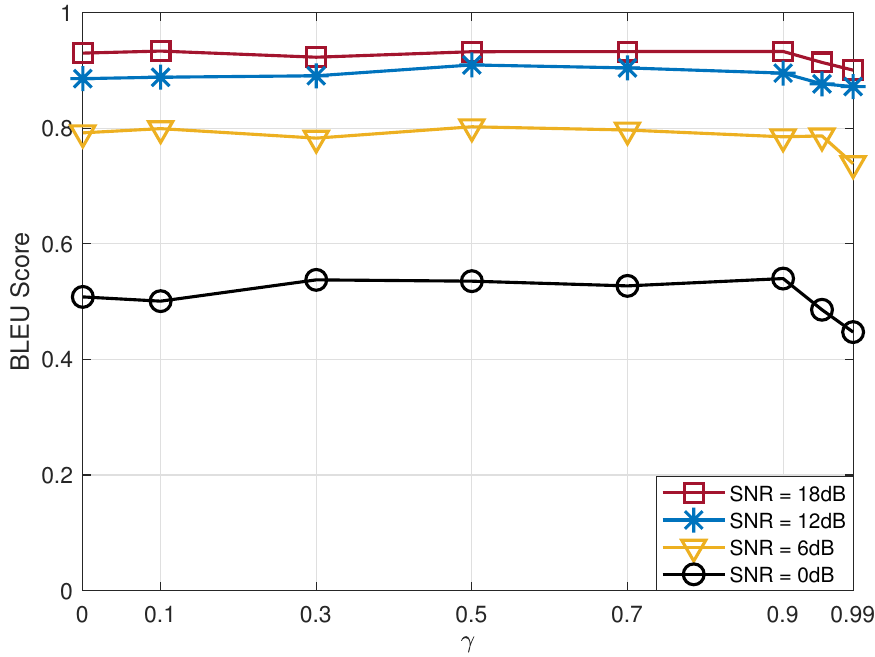}
	\caption{The BLEU scores of different SNRs versus sparsity ratio, $\gamma$, under Rician fadings channel with the refined CSI.}
	\label{prune1}
\end{figure}

Fig. \ref{prune1} shows the influences of network sparsity ratio, $\gamma$, on the BLEU scores with different SNRs under the Rician fading channels, where the system is pruned directly when  $\gamma$ increases from 0 to 0.9 and is pruned with fine-tuning when $\gamma$ increases to 0.99 continually. The proposed L-DeepSC achieves almost the same BLEU scores when the $\gamma$ increases from 0 to 0.9, which shows that there exists a mass of weights redundancy in the trained DeepSC model. When the $\gamma$ increases to 0.99, the BLEU scores still drop slightly due to the processing of fine-tuning, where the performance loss at 0 dB and 6 dB is larger than that at 12 dB and 18 dB. Thus, for the high SNR cases, the model can be pruned directly with only slight performance degradation. For the low SNR region, it is possible to prune 99\% weights without significant performance degradation when the system is sensitive to power consumption.

\begin{figure}[!t]
	\centering
	\includegraphics[width=85mm, height = 65mm]{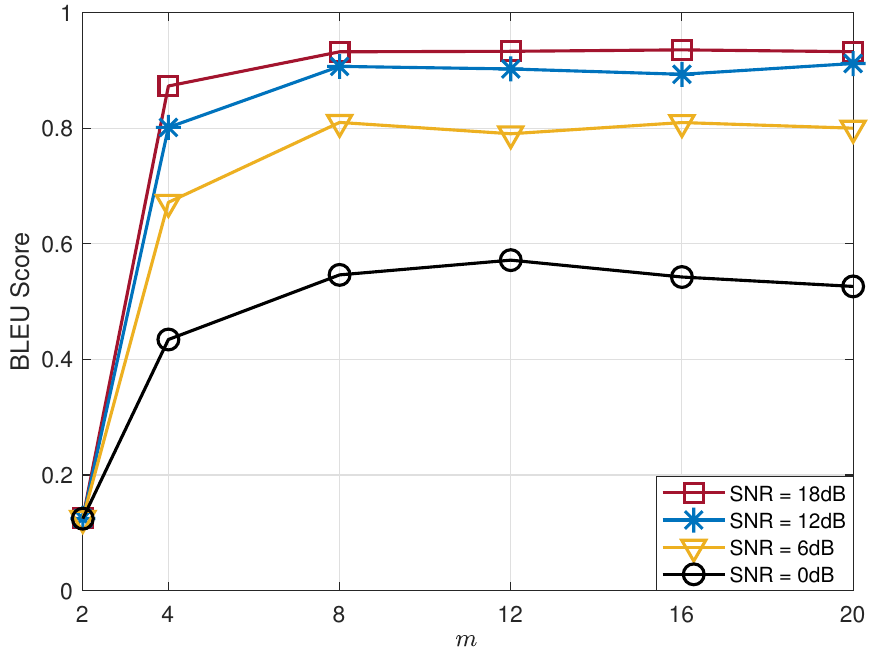}
	\caption{The BLEU scores of different SNRs versus quantization level, $m$, under Rician fading channels with the refined CSI.}
	\label{quantization}
\end{figure}

Fig. \ref{quantization} demonstrates the relationship between the BLEU score and the quantization bit number, $m$, under the Rician fading channels, where $m$ is defined in \eqref{eq21}, and the system is quantized with QAT when the $m$ is smaller than 2. The performance with $m=8$ to $m=20$ is similar, which indicates that the effectiveness of low-resolution neural networks. If the system is more sensitive to power consumption and can tolerant to certain performance degradation, the resolution of the neural networks can be further reduced to 4-bits level. However, the BLEU score decreases dramatically from $m=4$ to $m=2$ over the whole SRN range since most of the key information is removed in the low-resolution neural network.

\begin{table*}[tp!]
\footnotesize
\caption{The BLEU score and compression ratio, $\psi$, Comparisons versus different sparsity ratio, $\gamma$, and quantization level, $m$, in SNR = 12$dB$.}
\label{table 3}
\centering
\begin{tabular}{ |c |c c|c  c|c c|c c|c c|} 
\hline
\makecell[c]{Pruned Model} & \makecell[c]{BLEU score \\ with $m=4$} & \makecell[c]{$\psi$} &  \makecell[c]{BLEU score\\ with $m=8$} & \makecell[c]{$\psi$} &  \makecell[c]{BLEU score\\ with $m=12$} & \makecell[c]{$\psi$} &  \makecell[c]{BLEU score\\ with $m=16$} & \makecell[c]{$\psi$} &  \makecell[c]{BLEU score\\ with $m=32$} & \makecell[c]{$\psi$} \\
\hline
\makecell[c]{ $\gamma = 0$} & 0.811194 & 8 & 0.906763 & 4 & 0.902354 & 2.667 & 0.903089 & 2 & 0.895602 &1\\
\hline
\makecell[c]{ $\gamma = 0.3$} & 0.838967 & 11.429 & 0.892745 & 5.714 & 0.908537 & 3.81 & 0.910184 & 2.857 &0.89851 &1.429\\
\hline
\makecell[c]{ $\gamma = 0.6$} & 0.835863 & 20.0 & 0.897143 & 10.0 & 0.90815 & 6.667 & 0.900468 &5.0 &0.9093 & 2.5\\
\hline
\makecell[c]{ $\gamma = 0.9$} & 0.810322  & 80.0 & 0.895306 & 40.0 & 0.898784& 26.667 & 0.910554 & 20.0 & 0.89515 &10\\
\hline
\makecell[c]{ $\gamma = 0.95$} & 0.779685 & 160.0 & 0.875814 & 80.0 & 0.873426 & 53.333 & 0.877221 & 40.0 & 0.87653 & 20\\
\hline
\end{tabular}
\end{table*}

Table \ref{table 3} compares the BLEU scores and compression ratios under different combinations of weights pruning and weights quantization with SNR = 12 dB, where the compression ratio is computed by 
\begin{equation}\label{compression ratio}
\psi  = \frac{{M \times 32}}{{{M_{\tt pruned}} \times m}},
\end{equation}
where $M$ is the number of weights before pruning and $M_{\tt pruned}$ is the number of weights remaining after pruning,  32 is the number of required bits for FP32 and $m$ is the number of the required bits after quantization. The performance decreases when $\gamma$ increases or $m$ decreases, which are consistent with Fig. \ref{prune1} and Fig. \ref{quantization}. From the table, different compression ratios could lead to similar performance. For example, the BLEU score with $\gamma= 30 \%$ and $m=8$ is similar to that  with $\gamma =90\%$ and $m=12$, but the compression ratio is about five times different, i.e., 5.714 and 26.667. By properly choosing a suitable sparsity ratio and a quantization level, the same performance can be achieved but with a high compression ratio.

Table \ref{table 4} compares the DeepSC and L-DeepSC with 60\%  weights sparsity and 8-bit quantization when SNR is 12 dB, where we mainly consider the transmission of the weights. The simulation is performed in CPU by the  computer with Intel Core i7-9700CPU@3.00GHz.  After network slimmer, the model size is reduced from 12.3 MB to 1.28 MB while achieving a similar BLEU score, which means the bandwidth resource can be saved significantly without degrading the performance. Besides, the runtime slightly decreases from 20ms to 18ms since the unstructured pruning method is employed, and there exists the communication time between flash memory and some operation that can not be optimized. If the model size is bigger, the L-DeepSC could save more runtime.

\begin{table}[tp!]
\footnotesize
\caption{The comparison between L-DeepSC and DeepSC transceiver in parameters, size, runtime, and BLEU score.}
\label{table 4}
\centering
\begin{tabular}{ |c| c |c |c | c|} 
\hline
& Parameters &  Size &  Runtime & BLEU score \\
\hline
\makecell[c]{$\gamma = 0$, \\ $m = 32$} & 3,333,120  & 12.3 MB & 20ms & 0.895602\\
\hline
\makecell[c]{$\gamma = 0.6$, \\ $m = 8$} & 1,333,247 & 1.28 MB & 18ms & 0.897143\\
\hline
\end{tabular}
\end{table}

\section{Conclusion}
In this paper, we proposed a lite distributed semantic communication system, named L-DeepSC, for the Internet of Things (IoT) networks, where the participating devices are usually with limited power and computing capabilities. Specially, the receiver and feature extractor were designed jointly for text transmission. Firstly, we analyzed the effectiveness of CSI in forward-propagation and back-propagation during system training over the fading channels. The analytical results reveal that the fading channels contaminate the weights update and restrict model representation capability. Thus, a refined LS estimator with fewer pilot overheads was developed to eliminate the effects of fading channels. Besides, we map the full-resolution original constellation into finite bits constellation to lower the cost of IoT devices, which was verified by simulation results. Finally, due to the limited narrow bandwidth and computational capability in IoT networks, two model compression approaches have been proposed: 1) the network sparsification to prune the unnecessary weights, and 2) network quantization to reduce the weights resolution. The simulation results validated that the proposed L-DeepSC outperforms the traditional methods, especially in the low SNR regime, and has provided insights into the balance among compression ratio, sparsity ratio, and quantization level. Therefore,  our  proposed L-DeepSC is a promising candidate for intelligent IoT networks, especially in the low SNR regime.

% or


%\appendices

%\section{}

%\section*{Acknowledgment}


\ifCLASSOPTIONcaptionsoff
  \newpage
\fi



\bibliographystyle{IEEEtran}
\bibliography{reference.bib}
\end{document}